\documentclass[preprint]{iucr}              % DO NOT DELETE THIS LINE

\usepackage{multirow}
\usepackage{xcolor}

\begin{document}                  % DO NOT DELETE THIS LINE

     %-------------------------------------------------------------------------
     % The introductory (header) part of the paper
     %-------------------------------------------------------------------------

     % The title of the paper. Use \shorttitle to indicate an abbreviated title
     % for use in running heads (you will need to uncomment it).

\title{Van Vleck Analysis of Angularly Distorted Octahedra using VanVleckCalculator}
%\shorttitle{Short Title}

     % Authors' names and addresses. Use \cauthor for the main (contact) author.
     % Use \author for all other authors. Use \aff for authors' affiliations.
     % Use lower-case letters in square brackets to link authors to their
     % affiliations; if there is only one affiliation address, remove the [a].

\renewcommand*{\thefootnote}{\fnsymbol{footnote}}

\author{Liam. A. V.}{Nagle-Cocco\footnote{Email: lavn2@cam.ac.uk. ORCID: 0000-0001-9265-1588.}}{}{}%\footnote[1]{Email: lavn2@cam.ac.uk}
\author{Si\^an E.}{Dutton\footnote{Email: sed33@cam.ac.uk. ORCID: 0000-0003-0984-5504.}}{}{}

\aff{Cavendish Laboratory, University of Cambridge, JJ Thomson Avenue, Cambridge, CB3 0HE, \country{United Kingdom}}

\renewcommand*{\thefootnote}{\arabic{footnote}}
\setcounter{footnote}{0}

     % Use \shortauthor to indicate an abbreviated author list for use in
     % running heads (you will need to uncomment it).

%\shortauthor{Soape, Author and Doe}

     % Use \vita if required to give biographical details (for authors of
     % invited review papers only). Uncomment it.

%\vita{Author's biography}

     % Keywords (required for Journal of Synchrotron Radiation only)
     % Use the \keyword macro for each word or phrase, e.g. 
     % \keyword{X-ray diffraction}\keyword{muscle}

%\keyword{keyword}

     % PDB and NDB reference codes for structures referenced in the article and
     % deposited with the Protein Data Bank and Nucleic Acids Database (Acta
     % Crystallographica Section D). Repeat for each separate structure e.g
     % \PDBref[dethiobiotin synthetase]{1byi} \NDBref[d(G$_4$CGC$_4$)]{ad0002}

%\PDBref[optional name]{refcode}
%\NDBref[optional name]{refcode}

\maketitle                        % DO NOT DELETE THIS LINE

\begin{synopsis}
A method and associated Python script, \textsc{VanVleckCalculator}, is described for parametrising octahedral shear and first-order Jahn-Teller distortions in crystal structures.
\end{synopsis}

\begin{abstract}
Van Vleck modes \textcolor{black}{describe} all possible displacements of octahedrally-coordinated ligands about a core atom. They are a useful analytical tool for analysing the distortion of octahedra, particularly for the first-order Jahn-Teller distortion. %They are also effective for calculating octahedral shear, an approach rarely taken in the literature but which can yield useful information, as will be shown. 
    Determination of Van Vleck modes of an octahedron is complicated by the presence of angular distortion of octahedra however. This problem is most commonly resolved by calculating the bond distortion modes ($Q_2$, $Q_3$) along the bond axes of the octahedron, disregarding the angular distortion and losing information on the octahedral shear modes ($Q_4$, $Q_5$, and $Q_6$) in the process. In this paper, the validity of assuming bond lengths to be orthogonal in order to calculate the van Vleck modes is discussed, and a method is described for calculating Van Vleck modes without disregarding the angular distortion. A Python code for doing this, \textsc{VanVleckCalculator}, is introduced, and some examples of its use are given. Finally, we show that octahedral shear and angular distortion are often, but not always, correlated, and propose a parameter as the shear fraction, $\eta$. We demonstrate that $\eta$ can be used to predict whether the values will be correlated when varying a tuning parameter such as temperature or pressure.
\end{abstract}

     %-------------------------------------------------------------------------
     % The main body of the paper
     %-------------------------------------------------------------------------
     % Now enter the text of the document in multiple \section's, \subsection's
     % and \subsubsection's as required.

	\section{Introduction}

    The van Vleck distortion modes~\cite{van1939jahn} modes \textcolor{black}{describe} all possible displacements of octahedrally-coordinated ligands about a core atom. They are particularly useful in the context of the Jahn-Teller effect~\cite{jahn1937stability}, which in general occurs when a high-symmetry coordination is destabilised with respect to a deviation to lower symmetry as a consequence of electronic degeneracy. The Jahn-Teller effect distorts the crystal structure via the Jahn-Teller distortion. While the Jahn-Teller distortion is not unique to octahedra in bulk crystalline materials, it is in octahedra that it was first observed experimentally~\cite{bleaney1952cupric}, and it is in materials with Jahn-Teller-distorted octahedra that colossal magnetoresistance~\cite{millis1996dynamic} and high-temperature superconductivity~\cite{fil1992lattice,keller2008jahn} were discovered. 

\begin{figure}%[t]
    \includegraphics[scale=0.66]{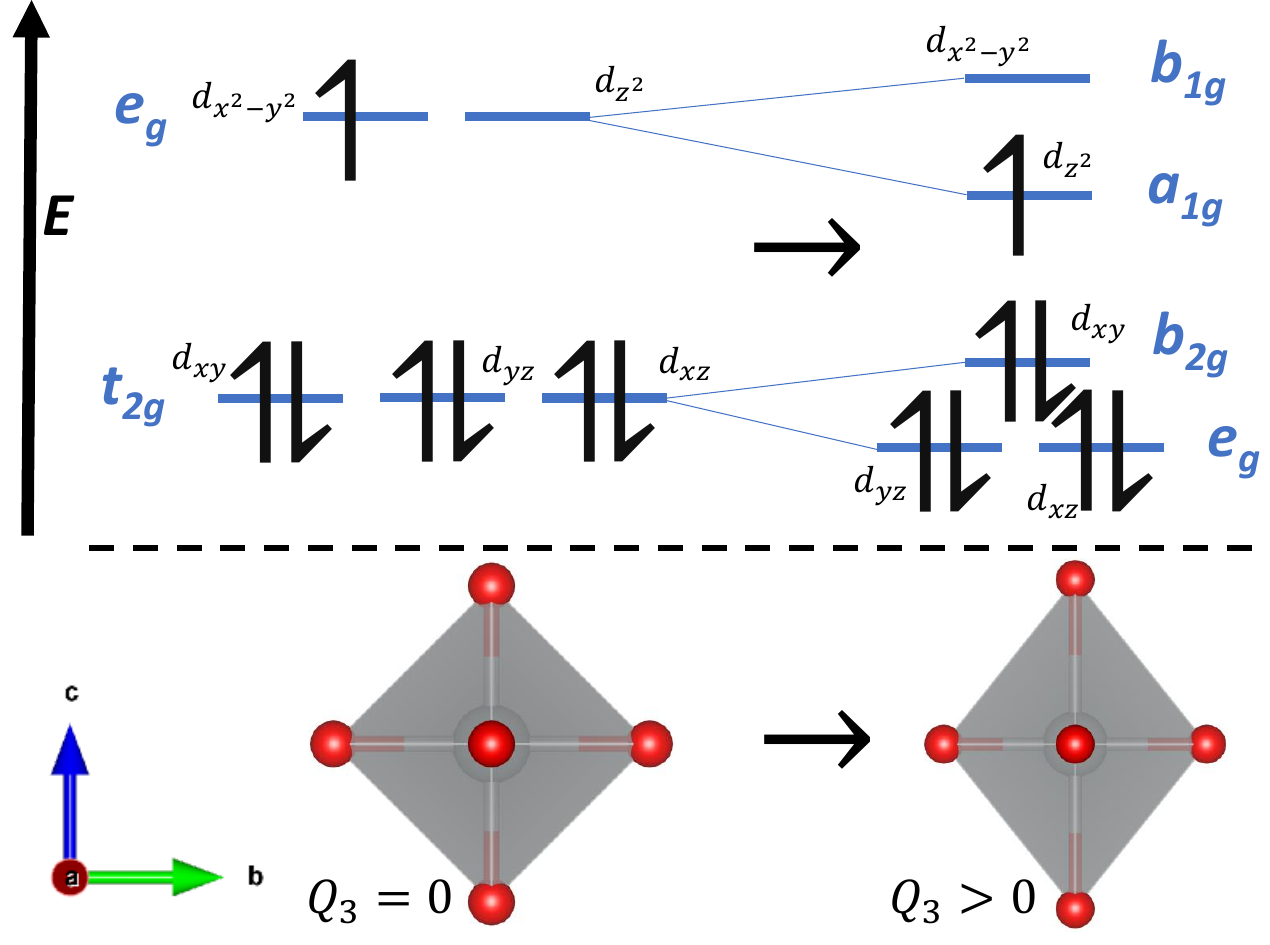}
    \caption{The orbital rearrangement due to a tetragonal elongation for an octahedrally-coordinated low-spin $d^7$ transition metal ion, which typically occurs due to the first-order Jahn-Teller effect.}
    \label{d_orbital_JT}
\end{figure}

     A transition metal (TM) cation in an octahedral configuration will have its $d$ orbitals split into three $t_{2g}$ orbitals\footnote{In this paper, we use the notation that lower case symmetry descriptors (such as $e_g$ or $t_{2g}$) refer to orbitals with this symmetry, and upper case descriptors (such as $E_g$ or $T_{2g}$) refer to the symmetry more generally.} at lower energy and two $e_g$ orbitals at higher energy. It will have a number, $n$, of electrons in these $d$ orbitals (hereafter described as $d^n$). For certain values of $n$ and, where applicable, certain low- or high-spin characters\footnote{In the low-spin case, $t_{2g}$ orbitals fill fully before $e_g$ orbitals gain electrons; in the high-spin case, once the $t_{2g}$ orbitals are singly-occupied, the next two electrons will populate the $e_g$ orbitals.}, there will exist multiple orbitals that could be occupied by an electron or an electron hole with equal energy. This degeneracy is destabilising, resulting in the most stable configuration of atomic sites being one in which the ligands distort from their high-symmetry positions in order to rearrange the orbitals into a non-degenerate system with minimised energy. This is shown for a low-spin $d^7$ TM cation (such as Ni$^{3+}$ or Co$^{2+}$) in Figure~\ref{d_orbital_JT}, though such distortions may occur for any value of $n$ in $d^n$ where there is a degenerate occupancy. The stabilisation energy due to the Jahn-Teller effect is larger for $e_g$ degeneracy than $t_{2g}$ degeneracy, and so the effect is prominent to higher temperatures, and hence more widely-studied, in JT-active materials with $e_g$ degeneracy~\cite{castillo2011highly}.

    In the literature, various techniques for parameterising the Jahn-Teller distortion are used. An often-used example~\textcolor{black}{\cite{kimber2012charge,lawler2021decoupling,nagle2022pressure,genreithschriever2023jahn}} is the bond length distortion index, defined by \citename{baur1974geometry} \citeyear{baur1974geometry}, as:

\begin{equation}\label{BLDI_equation}
	D=\frac{1}{n} \sum_{i=1}^n \frac{|l_i - l_\mathrm{av}|}{l_\mathrm{av}}
\end{equation}
	
	where $l_i$ is the distance between the core ion and the $i$th coordinated ion, and $l_\mathrm{av}$ is the average of all the distances between the core ion and coordinated ions.
    
    A similar parameter~\textcolor{black}{\cite{shirako2012high,sarkar2018rare,nagle2022pressure}} is the effective coordination number, which for an octahedron deviates from 6 only when there is bond length distortion, defined by \citename{hoppe1979effective} \citeyear{hoppe1979effective} as:
    
\begin{equation}\label{ECoN_equation}
	\mathrm{ECoN} = \sum_{i=1}^n \exp \left[ 1 - \left(\frac{l_i}{l'_\mathrm{av}} \right)^6 \right]
\end{equation}

where $l'_\mathrm{av}$ is a modified average distance defined as:

\begin{equation}
	l'_\mathrm{av} = 
	\frac{  \sum_{i=1}^n l_i \exp \left[ 1 - \left(\frac{l_i}{l_\mathrm{min}} \right)^6  \right]}	
	{ \sum_{i=1}^n   \exp \left[ 1 - \left(\frac{l_i}{l_\mathrm{min}} \right)^6 \right]}
\end{equation}
    
    Finally, a third parameter used to quantify the Jahn-Teller distortion~\cite{schofield1997distortion,kyono2015high,mikheykin2015cooperative} is the quadratic elongation, $<\lambda >$, defined by \citename{robinson1971quadratic} \citeyear{robinson1971quadratic} as:

\begin{equation}\label{QuadElon_equation}
<\lambda > = \frac{1}{n} \sum_{i=1}^n \left(\frac{l_i}{l_0}\right)^2
\end{equation}

where $l_0$ is the centre-to-vertex distance of a regular polyhedron of the same volume.

More recently, an alternative approach to modelling polyhedral distortion has been described~\cite{cumby2017ellipsoidal}, involving fitting an ellipsoid to the positions of the ligands around a coordination polyhedron, calculating the three \textcolor{black}{principal axes} of the ellipsoid, $R_1$, $R_2$, and $R_3$, where $R_1\le R_2 \le R_3$, and using the variance of these three radii as a metric for the distortion. This has been applied to the first-order Jahn-Teller distortion in \citename{pughe2023partitioning} \citeyear{pughe2023partitioning}.

These parameterisations each have merits. However, they are not sensitive to the symmetry of the octahedral distortion. The van Vleck modes are conceptually different to each of these for quantifying the Jahn-Teller distortion because they can be used to quantify distortion with the precise symmetry of the transition metal $e_g$ orbitals. This is important because Jahn-Teller distortions typically follow a particular symmetry. When the distortion is due to degeneracy in the $e_g$ orbitals it will be of $E_g$ symmetry; when it is due to degeneracy in the $t_{2g}$-degenerate orbitals it may be either $E_g$ or $T_{2g}$ symmetry~\textcolor{black}{\cite{child1965jahn,bacci1975coexistence,holland2002stereochemical,halcrow2009iron,teyssier2016jahn,schmitt2020electron,streltsov2022interplay}, although there is relatively little unambiguous experimental evidence for a Jahn-Teller-induced shear as compared with more typical $E_g$ distortion.}. 

    In this paper, we present a \textsc{Python}~\cite{10.5555/1593511} package, \textsc{VanVleckCalculator}, for calculating the van Vleck distortion modes. We show that the approach to calculating the modes which is commonly used in the literature is a reasonable approximation for octahedra with negligible angular distortion, but results in the loss of information in other cases. We propose a new metric, the shear fraction $\eta$, for understanding the correlation between octahedral shear and angular distortion. Finally, we re-analyse some previously-published data in terms of the van Vleck modes to show that these can be an effective way of understanding octahedral behaviour.

%    \subsection{Papers to cite}
% 
%	Ref.\cite{schmitt2020electron} - really nice discussion, relatively recent, lots of good references
%
%    Ref.\cite{van1939jahn} - the original paper
%    
%    Ref.\cite{khomskii2020orbital} - really useful review, see Fig 11 for instance
%
%    \cite{rodriguez1998neutron} - LaMnO3 variable temperature van vleck

%    Ref.~\cite{fedorova2018relationship} - a really nice paper, has van vleck modes and also discusses superexchange extensively

%    Ref.~\cite{khomskii2005role} - nice review by Khomskii, maybe not necessary to cite tho

\begin{figure}%[t]
    \includegraphics[scale=0.80]{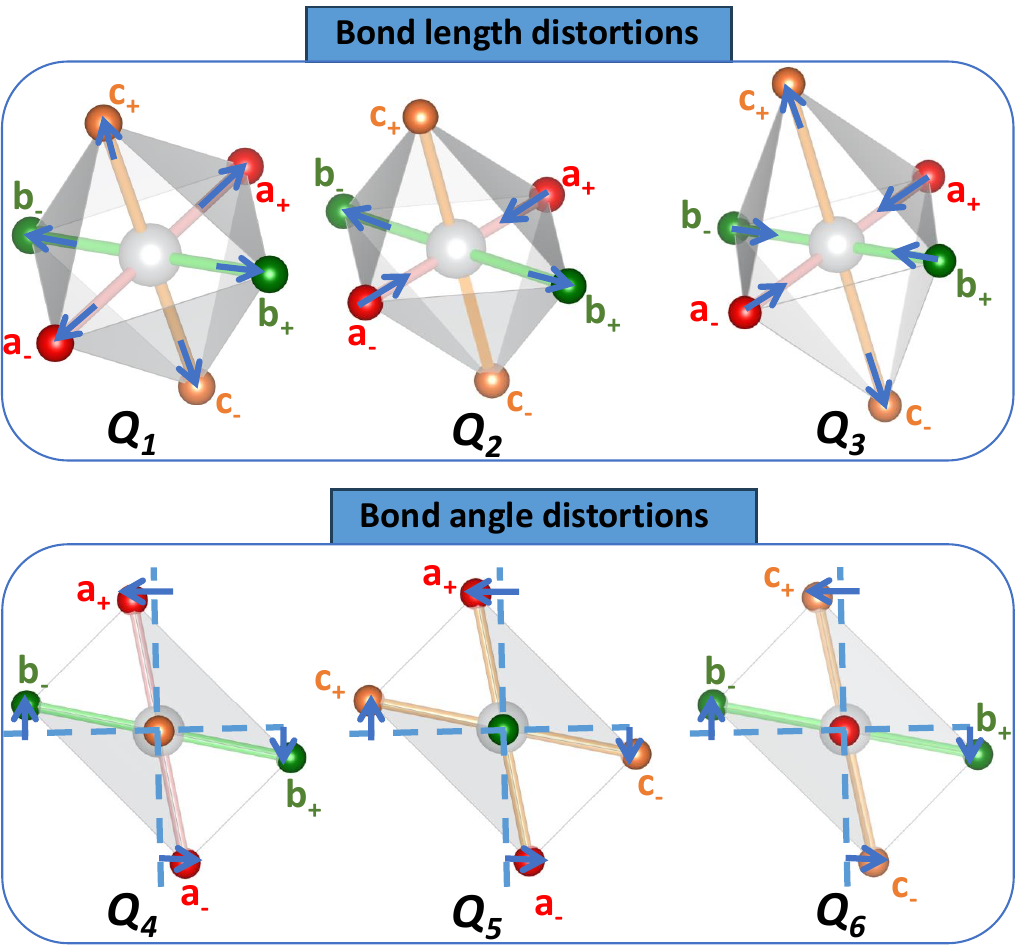}
    \caption{The 6 van Vleck modes exhibited for an octahedron, with sites labelled using the notation in the Theory section. For the octahedra exhibiting $Q_1$, $Q_2$, and $Q_3$ distortions, there is no angular distortion; for the octahedra exhibiting $Q_4$, $Q_5$, and $Q_6$ distortions, there is no bond length distortion. For the octahedral shear ($Q_4$, $Q_5$, and $Q_6$) modes, axes are drawn to show where the bond directions would be if undistorted. An octahedron can exhibit several, or all, of these distortions simultaneously.}
    \label{figure_showing_the_6_modes}
\end{figure}

    \section{Theory}

    Within an octahedron, we can split the 6 ligand ions into three pairs, where the two ions within the pair are opposite one another. In the absence of angular distortion (i.e., assuming all ligand-core-ligand angles are an integer number of 90$^\circ$), there would exist a basis where each of the three axes exist directly along the $x$-, $y$-, and $z$-axis, and where the origin in space is defined as the centre of the octahedron. 
    
    Each pair within an octahedron can therefore be assigned to an axis and labelled as the $a$, $b$, or $c$ pair respectively. Within a pair, ions can be labelled as $-$ or $+$ depending on whether they occur at a negative or positive displacement from the origin, along the axis, respectively. This notation is demonstrated in Figure~\ref{figure_showing_the_6_modes}, where each pair of ions is represented by a different colour.

    For each of the 6 ligands, we define a set of coordinates: $x^\alpha_\beta$, $y^\alpha_\beta$, and $z^\alpha_\beta$, where $\alpha$ is $a$, $b$, or $c$ denoting the pair in which the ligand is, and $\beta$ is $-$ or $+$ denoting which ion within the pair.

    The ideal positions of the six ligands are: ($R$,0,0), ($-R$,0,0), (0,$R$,0), (0,$-R$,0), (0,0,$R$), and (0,0,$-R$), where $R$ is defined as the distance between the centre of the octahedron and the ligand in an ideal octahedron (in practice, this is taken as the average of the core-ligand bond distances). This results in 18 independent \textcolor{black}{variables}. Using these, we further define a set of van Vleck coordinates (capitalised to distinguish from true coordinates) which is the displacement of the ion within an axis away from its ideal position. For instance, for the ion with $\alpha=a$ and $\beta=-$: $X^a_-$ = $x^a_- + R$, $Y^a_-$ = $y^a_-$, and $Z^a_-$ = $z^a_-$. See Figure~\ref{figure_showing_the_6_modes} for clarification of the ion notation.

    Using these coordinates, the first six van Vleck modes ($Q_j$; $j=1-6$) are defined as follows~\cite{van1939jahn}:

\begin{equation}\label{Q1_equation_definition}
    Q_1 = X^a_+ - X^a_- + Y^b_+ - Y^b_- + Z^c_+ - Z^c_-
\end{equation}

\begin{equation}\label{Q2_equation_definition}
    Q_2 = \frac{1}{2} \left[ X^a_+ - X^a_- - Y^b_+ + Y^b_- \right]
\end{equation}

\begin{equation} \label{Q3_equation_definition}
    Q_3 = \frac{1}{\sqrt{3}} \left[ \frac{1}{2} \left( X^a_+ - X^a_- + Y^b_+ - Y^b_- \right) - Z^c_+ + Z^c_- \right]
\end{equation}

\begin{equation}\label{Q4_equation_definition}
    Q_4 = \frac{1}{2} \left[ X^b_+ - X^b_- + Y^a_+ - Y^a_- \right]
\end{equation}

\begin{equation}\label{Q5_equation_definition}
    Q_5 = \frac{1}{2} \left[ Z^a_+ - Z^a_- + X^c_+ - X^c_- \right]
\end{equation}

\begin{equation}\label{Q6_equation_definition}
    Q_6 = \frac{1}{2} \left[ Y^c_+ - Y^c_- + Z^b_+ - Z^b_- \right]
\end{equation}

\begin{figure}%[t]
    \includegraphics[scale=1.1]{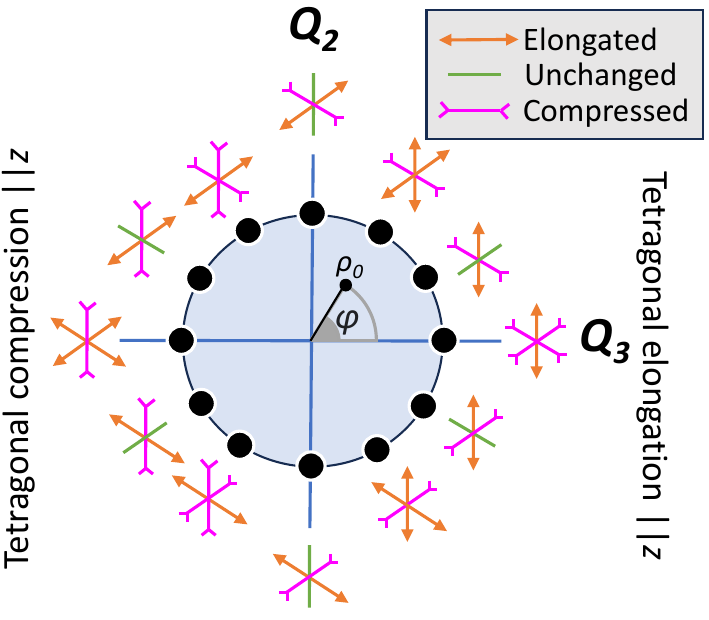}
    \caption{The $Q_2$-$Q_3$ phase space for elongated octahedra, with a representation of the values $\rho_0$ and $\phi$. Based on a figure from an article by \citename{Goodwin_two_temps} \citeyear{Goodwin_two_temps}.}
    \label{Q2_Q3_plane}
\end{figure}

We only discuss these first six van Vleck modes, which are shown in Figure~\ref{figure_showing_the_6_modes}. $Q_1$ to $Q_3$ describe bond length distortions, whereas $Q_4$ to $Q_6$ describe octahedral shear distortions. $Q_1$ is a simple expansion/contraction mode which does not affect symmetry and will not be discussed further.

$Q_2$ and $Q_3$ are a planar rhombic distortion and a tetragonal distortion respectively; they are considered degenerate due to the Hamiltonian, which is discussed for instance in \citename{kanamori1960crystal} \citeyear{kanamori1960crystal}. These two modes form a \textcolor{black}{basis for distortions describing different octahedral configurations} with the symmetry of the transition metal $e_g$ orbitals~\cite{goodenough1998jahn,khomskii2020orbital}. These modes are of most relevance for first-order Jahn-Teller distortions occurring due to degenerate $e_g$ orbitals. A phase space of possible octahedral configurations can be constructed using these two parameters~\cite{kanamori1960crystal}, as shown in Figure~\ref{Q2_Q3_plane}. Here the magnitude of the distortion $\rho_0$ can be calculated as follows:

\begin{equation} \label{rho_0}
    \rho_0 = \sqrt{{Q_2}^2+{Q_3}^2} % call it \rho_0 to be consistent with Goodenough and Zhou
\end{equation} 

and the angle\footnote{Note that this angle does not represent a physical angle within the octahedron.} $\phi$ of this distortion from being of purely $Q_3$ character can be calculated by:

\begin{equation}\label{phi_equation}
    \phi = \arctan{\left(\frac{Q_2}{Q_3}\right)}
\end{equation}

All possible combinations of the $Q_2$ and $Q_3$ modes correspond to a particular angle $\phi$, and hence a particular configuration as shown in Figure~\ref{Q2_Q3_plane}. The structural effect of a rotation of $\phi$ within a range of 120$^\circ$ can be quite significant, as shown in Figure~\ref{Q2_Q3_plane}; such changes can manifest as a Jahn-Teller-elongated\{compressed\} octahedron with 4 short\{long\} and 2 long\{short\} bonds (such as NiO$_6$ in NaNiO$_2$~\cite{nagle2022pressure}) or 2 short, 2 medium, and 2 long bonds (such as LaMnO$_3$~\cite{rodriguez1998neutron}). 

\begin{table}
  \caption{The special angles in the $Q_2$-$Q_3$ phase space [Figure~\ref{Q2_Q3_plane}], as a function of $\phi=\arctan\left(Q_2/Q_3\right)$, with the associated singly-occupied $e_g$ orbital, for $d^4$ and low-spin $d^7$ octahedral complexes. Note that for angles which are not special angles, there will be mixing of the orbital states of the nearest two special angles.}
  \label{table_special_angles}
  \begin{tabular}{ll}
    \hline
     $\phi$ ($^\circ$) & $\Psi(\phi)$ \\
    \hline
    0  & $d_{z^2}$\\
    60 & $d_{y^2-z^2}$\\
    120 & $d_{y^2}$\\
    180 & $d_{x^2-y^2}$\\
    240 & $d_{x^2}$\\
    300 & $d_{z^2-x^2}$\\
    \hline
  \end{tabular}
\end{table}

A characteristic of the Jahn-Teller distortion is that\textcolor{black}{, in the absence of external distortive forces,} the symmetry of the structure matches the symmetry of the orbitals involved. Typically, any $d$-orbital Jahn-Teller distortion will have some planar rhombic ($Q_2$) or tetragonal ($Q_3$) character. However, sometimes when the degeneracy occurs in the $t_{2g}$ orbital, there may instead be a trigonal component to the symmetry of the distortion, which manifests as an angular distortion instead~\cite{child1965jahn,bacci1975coexistence,holland2002stereochemical,halcrow2009iron,teyssier2016jahn,schmitt2020electron,streltsov2022interplay}. For the more commonly-studied case of a degeneracy in the $e_g$ orbitals, the effect of a rotation of $\phi$ similarly changes the symmetry of the $d$ orbitals. Figure~\ref{d_orbital_JT} shows the splitting of the $d$ orbitals in an octahedrally-coordinated $d^7$ transition metal due to an elongation-type first-order Jahn-Teller distortion, where the tetragonal elongation occurs along the $z$-axis. Note that the unpaired $e_g$ electron occupies the $d_{z^2}$ orbital. In the opposite case of a compression-type first-order Jahn-Teller distortion along the $z$ axis, the lower-energy, and hence singly-occupied, orbital would be the $d_{x^2-y^2}$; this would correspond to a rotation in $\phi$ of 180$^\circ$. More generally, as a function of $\phi$, there exist a set of special angles separated by a 60$^\circ$ rotation corresponding to a particular $e_g$ orbital being singly-occupied by a $d$ electron. These are tabulated in Table~\ref{table_special_angles}. An octahedron for which $\phi$ does not correspond to one of these special angles exhibits orbital mixing~\cite{rodriguez1998neutron,zhou2008orbital}. 

The $Q_4$ to $Q_6$ modes describe shear of the octahedra, i.e. the effect whereby paired ligands at opposite sides of a central ions are displaced in opposite directions%; this resembles the vibrational mode with $F_{2g}$ symmetry for octahedra~\cite{vtyurin2012hydrostatic}
, and have trigonal $T_{2g}$ \textcolor{black}{character}. The shear modes may be used to quantify the Jahn-Teller distortion in octahedra where the degeneracy occurs within $t_{2g}$ orbitals~\cite{child1965jahn,teyssier2016jahn}. The magnitude of the calculated shear is typically correlated with angular distortion, which is commonly quantified using the $\sigma_\zeta^2$ metric called the Bond Angle Variance~\cite{robinson1971quadratic} (BAV), defined here as:

\begin{equation}\label{BAV_equation}
\sigma^2_\zeta = \frac{1}{m-1} \sum_{i=1}^m (\zeta_i - \zeta_0)^2
\end{equation}

where $m$ is the number of bond angles (i.e. 12 for octahedra), $\zeta_i$ is the $i$th bond angle, and $\zeta_0$ is the ideal bond angle for a regular polyhedron (i.e. 90$^\circ$ for an octahedron). However, for direct comparison to the shear modes, it is more appropriate to use the standard deviation $\sigma_\zeta$.

\begin{figure}%[t]
    \includegraphics[scale=0.75]{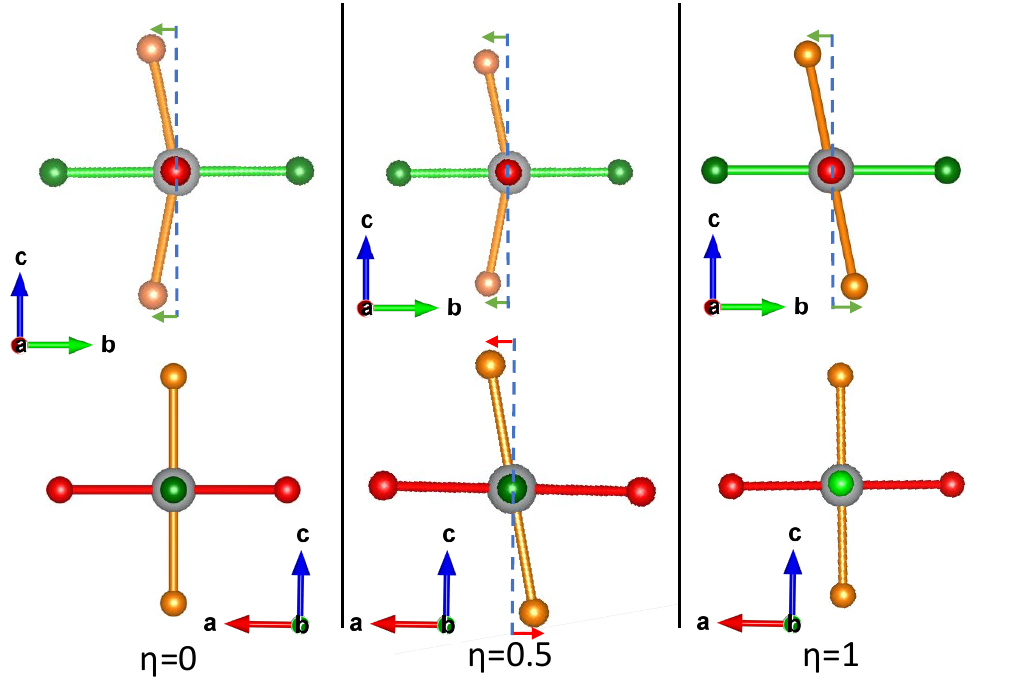}
    \caption{Three possible octahedral shear/anti-shear distortions, with the associated value of the shear fraction $\eta$ as defined in Equation~\ref{eta_equation}. In the case for $\eta=0$, the only distortion is anti-shear within a single plane. In the case for $\eta=0.5$, there are two planes in which there is distortion, a shear and anti-shear distortion equal in magnitude. In the case for $\eta=1$, there is a plane with a purely shear distortion.}
    \label{eta_figure}
\end{figure}

%It should be noted that while octahedral shear and angular distortion are typically correlated within a system, they may appear uncorrelated between systems. 
For an octahedron with non-zero $T_{2g}(Q_4,Q_5,Q_6)$ modes, increasing their magnitude will increase the angular distortion, but an octahedron may have angular distortion without exhibiting octahedral shear. To analyse the extent to which angular distortion in an octahedron is due to shear, we propose a shear fraction parameter $\eta$, demonstrated in Figure~\ref{eta_figure} and defined below.

First, we must define a set of shear and ``anti-shear" angular indices, which are modifications of Equations~\ref{Q4_equation_definition} to \ref{Q6_equation_definition} in terms of angles rather than displacements. The indices are represented with $\Delta$ and a subscript corresponding to the plane in which rotation occurs: the $ab$-plane corresponds to the $Q_4$ mode, the $ac$-plane to the $Q_5$ mode, and the $bc$-plane to the $Q_6$ mode. The absence or presence of a prime symbol, $\prime$, designates whether the index represents shear or anti-shear respectively. Finally, the $\delta$ angle is the rotation of the ligand from its ideal van Vleck coordinate in a clockwise direction, within the plane in which the corresponding van Vleck shear ($Q_4$ to $Q_6$) would occur. These are defined thus (see SI, Figure~S7):

\begin{equation}\label{Delta_ab_eq}
    \Delta_{ab} = \frac{1}{2} \left[ \delta^b_+ - \delta^b_- + \delta^a_+ - \delta^a_- \right]
\end{equation}

\begin{equation}\label{DeltaPrime_ab_eq}
    \Delta^\prime_{ab} = \frac{1}{2} \left[ \delta^b_+ + \delta^b_- - \delta^a_+ - \delta^a_- \right]
\end{equation}

\begin{equation}\label{Delta_ac_eq}
    \Delta_{ac}  = \frac{1}{2} \left[ \delta^a_+ - \delta^a_- + \delta^c_+ - \delta^c_- \right]
\end{equation}

\begin{equation}\label{DeltaPrime_ac_eq}
    \Delta^\prime_{ac}  = \frac{1}{2} \left[ \delta^a_+ + \delta^a_- - \delta^c_+ - \delta^c_- \right]
\end{equation}

\begin{equation}\label{Delta_bc_eq}
    \Delta_{bc}  = \frac{1}{2} \left[ \delta^c_+ - \delta^c_- + \delta^b_+ - \delta^b_- \right]
\end{equation}

\begin{equation}\label{Delta_prime_bc}
    \Delta^\prime_{bc}  = \frac{1}{2} \left[ \delta^c_+ + \delta^c_- - \delta^b_+ - \delta^b_- \right]
\end{equation}

We then quantify the shear and ``anti-shear" distortions using the following equations:

\begin{equation}\label{Deltamathrmshear2}
    \Delta_\mathrm{shear}^2 = \Delta_{ab}^2 + \Delta_{ac}^2 + \Delta_{bc}^2
\end{equation} 

\begin{equation}\label{Deltamathrmantishear2}
    \Delta_\mathrm{anti-shear}^2 = \Delta_{ab}^{\prime2} + \Delta_{ac}^{\prime2} + \Delta_{bc}^{\prime2}
\end{equation}

From here, we define the shear fraction $\eta$ as follows:

\begin{equation}\label{eta_equation}
    \eta  = \frac{\Delta_\mathrm{shear}^2}{\Delta_\mathrm{shear}^2+\Delta_\mathrm{anti-shear}^2}
\end{equation}

This $\eta$ parameter will be important in interpreting the relation between the angular distortion, $\sigma_\zeta$, and the van Vleck shear modes $Q_4$ to $Q_6$.

\section{Implementation}

In this section, the algorithm used to calculate Van Vleck distortion modes is discussed. It is written using \textsc{Python 3}~\cite{10.5555/1593511} as a
package called \textsc{VanVleckCalculator}, with the full code available on GitHub~\cite{VanVleckCalculator}, and also presented with annotations in the Supplementary Information. 
Data handling and some calculations make use of \textsc{NumPy}~\cite{harris2020array}, and crystal structures are handled using \textsc{PyMatGen}~\cite{Ong2012b}. 

A flow chart showing the octahedral rotation algorithm can be found in Supplementary Information, Figure~S1.

Besides calculating the van Vleck modes and the angular shear modes described in this paper, \textsc{VanVleckCalculator} can also calculate various other parameters as described in Supplementary Information.

\subsection{Selecting an origin}

Selection of the origin is a key step in calculating van Vleck modes. The most common approach, for an $M$X$_6$ octahedron, is to take the $M$ ion as the origin. This is a reasonable approach, given that $M$ ions are typically positioned at, or very close to, the centre of an octahedron. This is particularly appropriate for unit cells derived from Rietveld refinement~\cite{rietveld1969profile} of Bragg diffraction data, where the $M$ ion is likely to occur at a high-symmetry Wyckoff site. A third, similar, option would be to choose the average position of the 6 ligands as the origin in space. An example of when this may be a desirable choice would be for systems exhibiting a pseudo Jahn-Teller effect (also called the second-order Jahn-Teller effect), where the central cation is offset from the centre of the octahedron.

In some instances, a crystal structure may be simulated using a supercell. Examples include so-called ``big box" Pair Distribution Function (PDF) analysis~\cite{tucker2007rmcprofile} and Molecular Dynamics (MD)~\cite{bocharov2020ab} simulations. Such a supercell typically retains the periodicity which is an axiom of a typical crystallographic unit cell, but will exhibit local variations. For instance, a unit cell obtained by analysis of Bragg diffraction data is typically regarded as an ``average" structure, insensitive to local phenomena such as thermally-driven atomic motion or disordered atomic displacements such as a non-cooperative Jahn-Teller distortion. In a crystallographic unit cell, thermal motion of atoms is typically represented by variable Atomic Displacement Parameters (ADPs)~\cite{peterse1966anisotropic}. In contrast, a supercell should reflect local phenomena, for instance exhibiting local Jahn-Teller distortions in a system with a non-cooperative Jahn-Teller distortion, and representing thermal effects not with ADPs but rather by distributing equivalent atoms in adjacent repeating units in slightly different positions. In this regard, a supercell can be considered a ``snapshot" of a crystal system at a point in time. It may not be appropriate to set the core ion as the centre of the octahedron in a supercell, therefore, as the positioning of both core and ligand ions is in part due to thermal effects, and so the ``centre" of the octahedron will be displaced due to random motion. The alternative option would be to simply use the crystallographic site of the central ion and fix this as independent of the precise motion of the central ion locally. 

In \textsc{VanVleckCalculator}, the user has the option to take as the centre of the octahedron either the central ion, the average position of the 6 ligands, or a specified set of coordinates.

\subsection{Calculating van Vleck modes along bond directions}

The calculation of the van Vleck modes, as described in the Theory section, requires that the basis in space be the octahedral axes (i.e. the three orthogonal axes entering the octahedron via one vertex, passing through the central ion, and exiting via the opposite vertex). For a given crystal structure, this may require that an octahedron be rotated about each of the three axes making up the basis, until the octahedral axes perfectly align with the basis. This becomes more complicated when the octahedron exhibits angular distortion (i.e. exhibits ligand-core-ligand angles are not integer multiples of 90$^\circ$). In this case, it is impossible to define octahedral axes according to the strict criteria previously defined.

In the literature, this problem is generally evaded by simply calculating the Van Vleck modes on the basis of bond directions rather than Cartesian coordinates; for example, previous work on the perovskite LaMnO$_3$~\cite{goodenough1961relationship,rodriguez1998neutron,capone2000stabilization,chatterji2003volume,zhou2008orbital,zhou2011jahn,snamina2016spin,fedorova2018relationship,lindner2022interplay}, other perovskites~\cite{alonso2000evolution,wang2002intermediate,tachibana2007jahn,zhou2008intrinsic,castillo2011highly,franchini2011exceptionally,chiang2011effect,dong2012magnetic, fedorova2015biquadratic, ji2019strain,xu2020strategy,ren2021strain }, or non-perovskite materials~\cite{moron1993crystal,cussen2001control,wang2002role}.\footnote{We note that some works use a different variation which still uses Kanamori's approximation. Papers cited here include those which use the approximation, even if the precise definitions differ.} In this case, $Q_2$ and $Q_3$ are defined according to the following equations which were first expressed by \citename{kanamori1960crystal} \citeyear{kanamori1960crystal}, where $l$, $m$, and $s$ are the short, medium, and long bond lengths respectively\footnote{\textcolor{black}{The equations presented here differ from Kanamori's as they have been multiplied by a factor of $\frac{\sqrt{2}}{2}$, so that they are mathematically equivalent to Equations~\ref{Q2_equation_definition} and \ref{Q3_equation_definition}.}}:

\begin{equation}\label{schmitt_Q2}
    \textcolor{black}  {   Q_2 = l-s   }
\end{equation}

\begin{equation}\label{schmitt_Q3}
    \textcolor{black}  {   Q_3 = \frac{(2m-l-s)}{\sqrt{3}}   }
\end{equation}

This relies on the implicit assumption that bond lengths are orthogonal. This is clearly a reasonable approximation in many cases, particularly when angular distortion is very small. For instance, in LaMnO$_3$, the corner-sharing octahedral connectivity enables mismatched polyhedra to tessellate via octahedral tilting [Figure~\ref{LaAlO3_figure}(e)] rather than intra-octahedral angular distortion. However, for systems with greater angular distortion, for instance those with edge- or face-sharing interactions, it is not so clear that this approximation is valid. 

\subsection{Calculating van Vleck modes within Cartesian coordinates}

In \textsc{VanVleckCalculator}, we have written an algorithm for rotating an octahedron about three Cartesian axes with a defined origin within the octahedron, such that the ligands are as close as possible to the axes (within the constraint that there is angular distortion). This allows for calculation of van Vleck modes in a way that does not artificially constrain the octahedral shear modes ($Q_4$, $Q_5$, and $Q_6$) to be zero. 

First, three orthogonal axes are taken as the $x$-, $y$-, and $z$- axes\footnote{\textcolor{black}{We note that, for a set of three orthogonal vectors chosen as the axes, the choice to assign each to $x$, $y$, or $z$ will not affect the value of $\rho_0$, but will affect the value of $\phi=\arctan{\left(Q_2/Q_3\right)}$ by an integer multiple of 120$^\circ$, plus a reflection about the nearest special angle (see Table~\ref{table_special_angles}) if there is $Q_2$-$Q_3$ mixing.}}. By default, these are the [1,0,0], [0,1,0], and [0,0,1] axes respectively, but alternative sets of orthogonal vectors can be given by the user; for instance, for regular octahedra rotated 45$^\circ$ about the x axis, the user would be recommended to give as axes [1,0,0], [0,$\sqrt{2}$,$-\sqrt{2}$], and [0,$\sqrt{2}$,$\sqrt{2}$]. This vector is given as a \textsc{Python} list with shape (3,3). For consistency, the cross product of the first two axes should always be parallel with the third given vector; if anti-parallel, the algorithm will automatically multiply all elements in the third vector by -1. The three pairs of the octahedron (as defined in the Theory section) are each then assigned to one of these three axes on the basis of which pair has the largest projection of its displacement (the vector between two on a particular axis, with the $z$-axis assigned first, then the $y$-axis from amongst the two pairs not assigned to the $z$-axis, then the $x$ axis is automatically assigned to the remaining pair). Within each pair, the ligands are then ordered such that the ligand with the negative distance is along the assigned vector first, then the ligand with positive distance occurs second.

Second, the octahedron is rotated about the $x$-, $y$-, and $z$- directions of the basis repeatedly until the \textcolor{black}{orthogonal} axes supplied in the previous step match the basis precisely. This is performed in a while loop structure, with the rotation angles about the three axes summed in quadrature and compared with a defined tolerance (by default, $3\times10^{-4}$\,radians in \textsc{VanVleckCalculator}), and if the total rotation exceeds the tolerance, the step is repeated\footnote{This is because rotation operations do not commute, and so a single rotation about each axis \textcolor{black}{is unlikely to} result in the defined axes being superimposed over the basis vectors.}. This step is usually unnecessary, and can be skipped by leaving the default set of orthogonal axes, which are [1,0,0], [0,1,0], and [0,0,1] (meaning no rotation will occur). 

Third, an automatic rotation algorithm will further minimise the effect of angular distortion. For each of the three axes, the four ligands not intended to align with that axis are selected. The angle to rotate these four ligands about the origin such that each is aligned with its intended axis within the plane perpendicular to the axis of rotation is calculated. The octahedron is then rotated about this axis by the average of these four angles. This occurs iteratively until, for a given iteration, the sum (in quadrature) of the three rotation angles is less than the already-mentioned defined tolerance.

At this point, the octahedron is \textcolor{black}{optimally aligned with the basis (given the limitation that there may be angular distortion) }and the van Vleck modes can be calculated.

\subsection{Ignoring or including angular distortion: a comparison}

To evaluate the utility of calculating the van Vleck modes without disregarding the angular distortion, we perform a comparison between the two approaches. We have calculated the van Vleck distortion modes and associated parameters for octahedra in NaNiO$_2$ and LaMnO$_3$ with both a method that ignores angular distortion and calculates modes along bond directions (consistent with the $Q_2$ and $Q_3$ equations defined by \citename{kanamori1960crystal} \citeyear{kanamori1960crystal}), and a method that used Cartesian coordinates in order to take angular distortion into account. Table~\ref{table_angular-distortion_LaMnO3-NaNiO2} shows this for these two materials. Firstly, \textcolor{black}{for the van Vleck modes calculated} without ignoring angular distortion, we can see the octahedral shear modes ($Q_4$, $Q_5$, $Q_6$) are larger for the material with higher angular distortion \textcolor{black}{(as quantified using bond angle variance)}. While the effect of ignoring angular distortion is significant for the $Q_4$, $Q_5$, and $Q_6$ modes, it makes negligible difference for the calculation of $Q_2$ and $Q_3$ modes, and the associated $\rho_0$ and $\phi$ parameters. It is therefore likely a reasonable approximation to take, particularly for calculation of $\phi$ as is common in literature, even for octahedra which exhibit higher angular distortion. However, there is a definite loss of information in assuming the shear modes $Q_4$ to $Q_6$ are zero. The impact of this is assessed in the case studies.

\begin{table}%[t]
  \caption{A comparison between calculated $\phi$, $\rho_0$, $Q_2$, $Q_3$, $Q_4$, $Q_5$, $Q_6$, and $\eta$ values for NaNiO$_2$ and LaMnO$_3$ at room-temperature, calculated using orthogonal axes as described in this report (``Cartesian" method), and alternatively by ignoring angular distortion and calculating van Vleck modes along bond lengths (``Kanamori" method\footnote{So named because the equations originate in \citename{kanamori1960crystal} \citeyear{kanamori1960crystal}}). The centre of the octahedron is taken as the central ion position. Values were calculated using crystal structures reported on the Inorganic Crystal Structure Database (ICSD). To demonstrate the difference in angular distortion, the Bond Angle Variance (defined in Equation~\ref{BAV_equation}) is also tabulated. BAV is rounded to the third significant figure; $Q$ modes and related parameters are rounded to the 4th decimal place.}
  \label{table_angular-distortion_LaMnO3-NaNiO2}
  \begin{tabular}{c||c|c||c|c||c|c}
    \hline
    & \multicolumn{2}{c}{NaNiO$_2$} & \multicolumn{2}{c}{LaMnO$_3$} \\
    \hline
     ICSD code &  \multicolumn{2}{c}{415072} & \multicolumn{2}{c}{50334} \\
     Ref. &  \multicolumn{2}{c}{\citename{sofin2005new} \citeyear{sofin2005new}} & \multicolumn{2}{c}{\citename{rodriguez1998neutron} \citeyear{rodriguez1998neutron}} \\
     Octahedron &  \multicolumn{2}{c}{NiO$_6$} & \multicolumn{2}{c}{MnO$_6$} \\
     JT-active &  \multicolumn{2}{c}{Yes} & \multicolumn{2}{c}{Yes} \\
     Connectivity & \multicolumn{2}{c}{edge}  & \multicolumn{2}{c}{corner} \\
     BAV ($^{\circ2}$) & \multicolumn{2}{c}{35.2}  & \multicolumn{2}{c}{0.45} \\
     \hline
    Method & Kanamori  & Cartesian & Kanamori & Cartesian \\
%     $Q_1$ & 0.0126 & 0.0000 & 0.0002 & 0.0000 \\
     $Q_2$ (\AA{}) & 0.0000 & 0.0000 & 0.2745 & 0.2745 \\
     $Q_3$ (\AA{}) & 0.2834 & 0.2833 & -0.0860 & -0.0860 \\
     $Q_4$ (\AA{}) & 0 & 0.2078 & 0 & 0.0130 \\
     $Q_5$ (\AA{}) & 0 & 0.2001 & 0 & 0.0114 \\
     $Q_6$ (\AA{}) & 0 & 0.2001 & 0 & 0.0361 \\
     $\phi$ ($^\circ$) & 0.0000\footnote{Note that $\phi=0^\circ$ is equivalent to 120$^\circ$ or 240$^\circ$.} & 0.0000 & 107.3929 & 107.4034 \\
     $\rho_0$ (\AA{}) & 0.2834 & 0.2833 & 0.2877 & 0.2876 \\
     $\Delta_\mathrm{shear}$ (\AA{}) & N/A & 0.3534 & N/A & 0.0389 \\
     $\Delta_\mathrm{anti-shear}$ (\AA{}) & N/A & 0 & N/A & 0 \\
     $\eta$ & N/A  & 1.0 & N/A & 1.0 \\
    \hline
  \end{tabular}
\end{table}

\section{Case studies}

\subsection{Temperature-dependence of octahedral shear in LaAlO$_3$}

Perovskite and perovskite-like crystal structures are amongst the most important and widely-studied crystalline material classes in materials science today. Perovskite crystal structures have \textit{ABX}$_3$ chemical formulae, with \textit{A} and \textit{B} being ions at the centres of dodecagons and octahedra, respectively, with the \textit{X} anion constituting the vertices of these polyhedra. The \textit{BX}$_6$ octahedra interact via corner-sharing interactions. There are also perovskite-like crystal structures such as the double perovskites, $A_2BB^\prime X_6$~\cite{king2010cation,koskelo2023magnetic}, for which many of the same principles apply. 

\begin{figure}%[t]
    \includegraphics[scale=0.85]{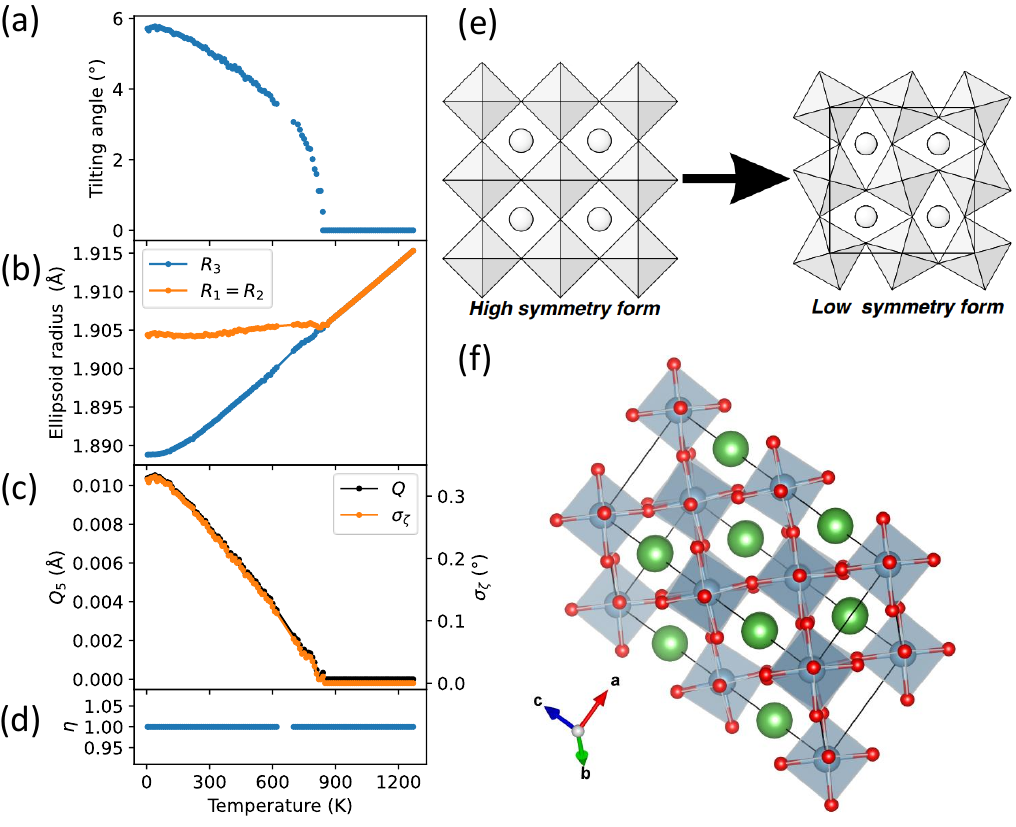}
    \caption{The results of our analysis on LaAlO$_3$ as a function of temperature. (a) octahedral tilting angle as reported by \textcolor{black}{\citename{hayward2005transformation} \citeyear{hayward2005transformation}} and extracted using DataThief III~\cite{Tummers_2006}. (b) radii of the minimum bounding ellipsoid fitted to the O anions of the AlO$_6$ octahedra using \textsc{PIEFACE}~\cite{cumby2017ellipsoidal}. (c) octahedral shear parameter $Q_5$ of the AlO$_6$ octahedra, where $Q_5=-Q_4=-Q_6$, calculated using \textsc{VanVleckCalculator}, compared with $\sigma_\zeta$ the bond angle standard deviation (orange). (d) shear fraction $\eta$, defined in Equation~\ref{eta_equation}. (e) the transition between low-symmetry (tilting) and high-symmetry (tilt-free) perovskite structures, adapted with permission from \citename{angel2005general}, APS Physical Review Letters, 95, 025503, copyright 2005 American Physical Society. (f) the perovskite crystal structure of LaAlO$_3$ at 4.2\,K from \citename{hayward2005transformation} \citeyear{hayward2005transformation}.}
    \label{LaAlO3_figure}
\end{figure}

The ideal perovskite system would be cubic, with space group $Pm\bar{3}m$, but many related structures with lower symmetry are known. This typically occurs in three situations~\cite{woodward1997octahedral}:
\begin{enumerate}
    \item when there is a mismatch between the ionic radii of the octahedrally-coordinated \textit{BX}$_6$ cation and the dodecagonally-coordinated \textit{AX}$_{12}$ cation, resulting in tilting of the octahedra; see Figure~\ref{LaAlO3_figure}(e).
    \item when there is displacement of the central cation from the centre of the octahedron, typically due to the pseudo Jahn-Teller effect.
    \item when the ligands of the octahedron are distorted by electronic phenomena such as the first-order Jahn-Teller effect.
\end{enumerate}

In this case study, we focus on the first case, where a size mismatch results in octahedral tilting. Octahedra are often modelled as rigid bodies, but in practice they are not rigid in all systems, and the octahedral tilting will often induce strain resulting in angular distortion. This is typically far smaller than that seen in edge-sharing materials such as NaNiO$_2$, but it is large enough that it cannot be disregarded when attempting to fully understand the structure of the material. As was noted by Darlington~\cite{darlington1996phenomenological}, this angular distortion commonly manifests as shear.

LaAlO$_3$ is a perovskite-like $ABX_3$ material which is cubic (space group $Pm\bar{3}m$) above around $\sim$830\,K, but which exhibits a rhombohedral distortion below this temperature (with space group $R\bar{3}c$) due to octahedral tilting~\cite{hayward2005transformation}, see see Figure~\ref{LaAlO3_figure}(e) and (f). Throughout both temperature regimes, there is the absence of bond length distortion; a calculation of the bond length distortion index would yield a value of zero at all temperatures. In the low-temperature regime, the magnitude of the distortion continuously decreases with increasing temperature, reaching zero at the transition temperature. Most commonly in the literature, the tilting angle between the octahedral axis and the $c$-axis (0$^\circ$ in the cubic phase) is used to quantify this distortion; for LaAlO$_3$, this is shown in Figure~\ref{LaAlO3_figure}(a). The strain induced by this distortion results in intra-octahedral angular distortion. \citename{hayward2005transformation} \citeyear{hayward2005transformation} model this in terms of strain tensors, finding a linear temperature dependence below the transition temperature, which differs from the temperature-dependence of the tilting angle (which resembles an exponential decline), implying the two are related but distinct phenomena. \textcolor{black}{\citename{cumby2017ellipsoidal} \citeyear{cumby2017ellipsoidal}} instead model the octahedral distortion for this same dataset using the radii of a minimum-bounding ellipsoid, and also find approximately linear temperature dependence of the long and short radii as they approach convergence (see Figure~\ref{LaAlO3_figure}(b)). 

Here, we calculate the van Vleck shear modes. Due to the symmetry of the octahedral tilting, there is only one independent shear mode, and $Q_5=-Q_4=-Q_6$. We compare this with the bond angle standard deviation given in Equation~\ref{BAV_equation}, see Figure~\ref{LaAlO3_figure}. We see that despite being distinct parameters, the temperature dependence of both is entirely identical. We attribute this to the shear fraction, $\eta$, being precisely 1 for all temperatures where there is angular distortion, meaning that shear is completely correlated with angular distortion. 

%Note: the following line was commented out on 9th July 2023 as I move this case study to occur first in the paper.
%While this behaviour can be considered to differ with the pressure-dependence of NaNiO$_2$ shown in Figure~\ref{VanVleck_pressure2} despite both having $\eta\approx1$, this is simply because NiO$_6$ octahedra in NaNiO$_2$ have two unique shear modes whereas AlO$_6$ octahedra in LaAlO$_3$ have only one.

% Note to self: eta in Hayward's paper is not really relevant here. It can be found defined in Christopher J Howard's 2000 paper "Neutron powder diffraction study of rhombohedral rare-earth aluminates and the rhombohedral to cubic phase transition", equation 2

\subsection{Big box analysis of Pair Distribution Function data on LaMnO$_3$}

\begin{figure}%[t]
    \includegraphics[scale=0.9]{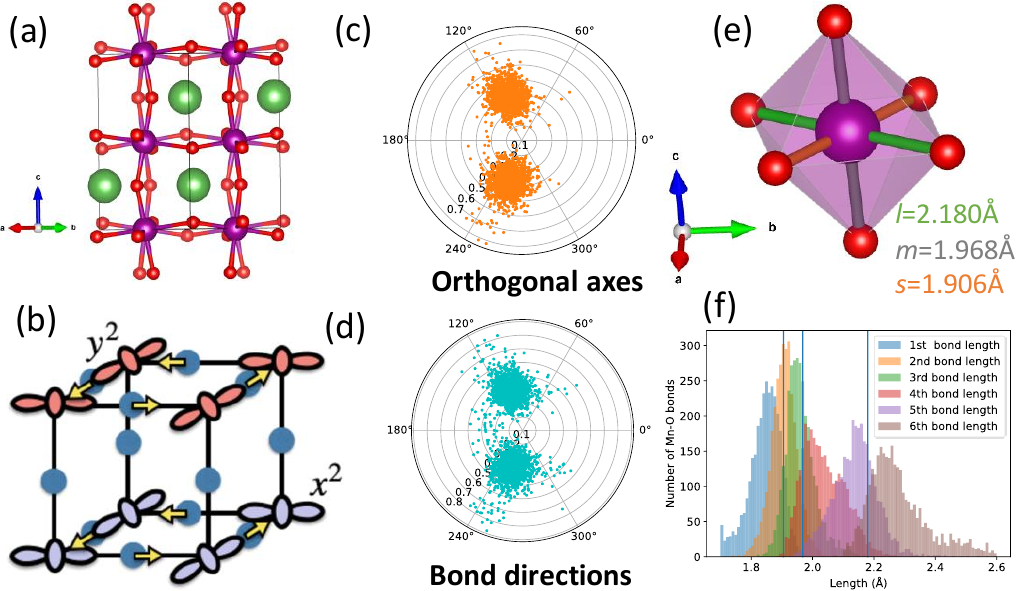}
    \caption{The perovskite-like structure of LaMnO$_3$, as obtained from ICSD structure 50334, is shown in (a). (b) the orbital ordering at room-temperature in LaMnO$_3$, and is reprinted with permission from \citename{khomskii2020orbital}, Chem. Rev. (2021), 121, 5, 2992–3030, copyright 2021 American Chemical Society. (c) and (d) show polar plots with a point representing the calculated $\phi$ and $\rho_0$ values for each MnO$_6$ octahedron in a $10\times10\times8$ supercell of LaMnO$_3$ at room-temperature, as obtained from reverse Monte Carlo analysis of neutron Pair Distribution Function data in \citename{thygesen2017local} \citeyear{thygesen2017local}. In (c), orthogonal axes were used (i.e. angular distortion was included in the calculation, using the method described in this manuscript), whereas in (d) the Mn-O bond directions were taken as the axes regardless of orthogonality. (e) the Mn$^{3+}$ octahedra which exhibit a mixed $Q_2$-$Q_3$ type distortion due to the first-order Jahn-Teller effect, manifesting as three different bond lengths, labelled in ascending order of length as $s$ (orange), $m$ (grey), and $l$ (green). (f) a histogram of the smallest to largest Mn-O bond length within each octahedron in the $10\times10\times8$ supercell, with the blue vertical lines indicating the bond lengths in the average structure.}
    \label{LMO-fig}
\end{figure}

The Jahn-Teller distortion in LaMnO$_3$, a perovskite-like $ABX_3$ material which has the crystal structure shown in Figure~\ref{LMO-fig}(a), occurs as a consequence of degeneracy in the $e_g$ orbitals on the high-spin $d^4$ Mn$^{3+}$ ion. At ambient temperatures, it is a prime example of a cooperative Jahn-Teller distortion, exhibiting long-range orbital order where the elongation of the Jahn-Teller axis alternates between the $a$ and $b$ directions for neighbouring MnO$_6$ octahedra, never occurring along the $c$ direction~\cite{khomskii2020orbital} [Figure~\ref{LMO-fig}(b)]. With heating through $\sim$750\,K, the Jahn-Teller distortion can no longer be observed in the average structure obtained from Bragg diffraction~\cite{rodriguez1998neutron}. However, the Jahn-Teller distortion persists locally as has been shown by pair distribution function~\cite{qiu2005orbital} and EXAFS~\cite{garcia2005jahn,souza2005local} measurements. This transition is one of the most widely-studied orbital order-disorder transitions for the first-order Jahn-Teller distortion. The high-temperature orbital regime has been described theoretically in terms of a three-state Potts model~\cite{ahmed2006potts,ahmed2009volume}, a view supported by big box analysis of combined neutron and x-ray pair distribution function data~\cite{thygesen2017local}, as performed using RMCProfile~\cite{tucker2007rmcprofile}.

In this case study, we take a $10\times10\times8$ supercell of LaMnO$_3$, obtained using RMCProfile against total scattering data obtained at room-temperature, and previously published in the aforementioned work~\cite{thygesen2017local}. Results are shown in Figure~\ref{LMO-fig}. We repeat the analysis of this supercell from the perspective of the $E_g(Q_2,Q_3)$ van Vleck distortion modes, using two different approaches: (1) the algorithm for automatically determining a set of orthogonal axes is applied to each octahedron individually%
%, (2) for each octahedron the van Vleck modes are calculated using bond lengths projected onto the same set of orthogonal axes\footnote{These are $[ 3 , 2 , -0.5]$, $[-2.25 , 3.5 , 0.5]$, and $[ 2.75 , -0.375, 15.   ]$, or an analogous set of axes rotated about the $ab$ plane to account for octahedral tilting.}, and (3)
, and (2) following the van Vleck equations~\ref{schmitt_Q2} and \ref{schmitt_Q3} proposed by \citename{kanamori1960crystal} \citeyear{kanamori1960crystal} where angular distortion is disregarded. In each of these cases the crystallographic site of the supercell is taken as the origin, and so thermally-driven variations in the Mn position will not affect the result.

As can be seen in Figures~\ref{LMO-fig}(d)-(f), there are two clusters of octahedra within the polar plot, occurring at $\phi\approx\pm107^\circ$. \textcolor{black}{This corresponds to occupation of the $d_{y^2}$ orbitals ($+$) and of the $d_{x^2}$ orbitals ($-$).} In both cases, the superposition of perpendicular $Q_3$ compression and elongation modes results in an octahedron with mixed $Q_2$-$Q_3$ character. This finding is consistent with previous works which placed MnO$_6$ octahedra from LaMnO$_3$ into the framework of an $E_g(Q_2,Q_3)$ polar plot~\cite{zhou2008intrinsic,zhou2011jahn}.

Figure~\ref{LMO-fig}(e) shows the MnO$_6$ octahedron in the average structure of LaMnO$_3$ at room temperature, with the three different bond lengths plotted in Figure~\ref{LMO-fig}(f) along with a histogram of all the bond lengths in the supercell. This shows how the combination of the $Q_2$ and $Q_3$ distortion modes manifests in the octahedral distortion. 
%The reason for the $Q_2$ component cannot be due to an increased stability from rearranging the Mn $3d$ orbitals, for the following reasons. Firstly, as was shown in Figure~\ref{d_orbital_JT}, a purely $Q_3$ elongation along the $z$ direction makes the $t_{2g}$ orbitals degenerate between $\ket{d_{yz}}$ and $\ket{d_{xz}}$, with $\ket{d_{xy}}$ at a higher energy. A $Q_2$-type distortion breaks the degeneracy between the $\ket{d_{yz}}$ and $\ket{d_{xz}}$ orbitals~\cite{khomskii2020orbital}. However, for $d^4$ Mn$^{3+}$ the $t_{2g}$ orbitals are singly-occupied, and so the total energy is not reduced by a $Q_2$ component to the distortion. Secondly, in edge-sharing HS $d^4$ Mn$^{3+}$ compounds such as $\alpha$-NaMnO$_2$ (checked using ICSD references 15769~\cite{jansen1973kenntnis} and 21028~\cite{hoppe1975kenntnis}) and LiMnO$_2$ (checked using ICSD reference 82993~\cite{armstrong1996synthesis}), there is no $Q_2$ component to the Jahn-Teller-distorted octahedra, which there would be if energy could be minimised via such a distortion. 

\textcolor{black}{The $Q_2$ contribution to the distortion, as seen from the three different Mn-O bond lengths in LaMnO$_3$, is also present in Jahn-Teller-distorted $A$CuF$_3$ ($A$=Na,K,Rb)~\cite{lufaso2004jahn,marshall2013unusual,khomskii2020orbital} and even in some Jahn-Teller-undistorted perovskites~\cite{zhou2008intrinsic}, indicating it is related to the structure. It is not intrinsic to Jahn-Teller-distorted manganates, as it is absent in high-spin $d^4$ Mn$^{3+}$ with edge-sharing octahedral interactions and colinear orbital ordering such as $\alpha$-NaMnO$_2$ and LiMnO$_2$ (checked using ICSD references 15769 and 82993~\cite{jansen1973kenntnis,armstrong1996synthesis} respectively). The $Q_2$ component to the octahedral distortion is therefore likely intrinsic to the crystal structure~\cite{zhou2006unusual,zhou2008intrinsic}, which occurs as a result of octahedral tilting reducing the symmetry from cubic $Pm\bar{3}m$ to $Pnma$. In LaMnO$_3$, the combination of the $Q_2$ component to the distortion and the orbital ordering [Figure~\ref{LMO-fig}(b)] are a possible distortion of the $Pnma$ space group. In this way, the orbital ordering may be coupled to the octahedral tilting, a link previously made by \citename{lufaso2004jahn} \citeyear{lufaso2004jahn}.}

Finally, we also calculate the $Q_4$ to $Q_6$ octahedral shear modes for all octahedra in the supercell, presented as a histogram in Figure~S2 %\ref{LMO_shear-SI} 
in Supplementary Information. We present the average and standard deviation, as calculated assuming orthogonal axes and with the automated octahedral rotation: $Q_4=-0.02\pm0.13$\,\AA{}, $Q_5=0.02\pm0.10$\,\AA{}, and $Q_6=-0.00\pm0.11$\,\AA{}. In each case, the magnitude of the distortion is zero within standard deviation, and also contains the value from the average structure presented in Table~\ref{table_angular-distortion_LaMnO3-NaNiO2} within the range of error. This low level of shear generally supports the validity of calculating the $E_g(Q_2,Q_3)$ van Vleck modes along bond directions rather than a Cartesian coordinate system for a system like LaMnO$_3$. %However, it is likely the consequence of the supercell PDF analysis method simulating thermal effects by inducing atomic site disorder, rather than indicating the absence of shear, because the strain from octahedral tilting will induce shear, like for LaAlO$_3$. 
It is interesting to note that the standard deviation is higher for $Q_4$, which quantifies the shear within the plane in which there is orbital ordering.

\subsection{Effect of pressure on the JT distortion in NaNiO$_2$}

In recent years, there have been several studies looking at the effect of applied pressure on the Jahn-Teller distortion in crystalline materials~\cite{aasbrink1999high,loa2001pressure,choi2006electronic,zhou2008breakdown,zhou2011jahn,aguado2012pressure,mota2014dynamic,caslin2016competing,zhao2016pressure,collings2018disorder,bhadram2021pressure,lawler2021decoupling,scatena2021pressure,ovsyannikov2021structural,nagle2022pressure}. Most of these have shown that, as a general rule, pressure reduces the magnitude of the Jahn-Teller distortion as a consequence of the elongated bond being more compressible than the shorter bonds. 
%
%Various parameters have been used to quantify this change with pressure, such as bond length distortion index (defined in Ref.~\cite{baur1974geometry}). However, the parameters most specialised in application to the Jahn-Teller distortion, the van Vleck modes, are used relatively rarely. 
\citename{zhou2011jahn} \citeyear{zhou2011jahn} use van Vleck modes to quantify the effect of pressure on the Jahn-Teller distortion in the corner-sharing perovskite-like compounds LaMnO$_3$ and KCuF$_3$. While application of pressure reduces the magnitude of the distortion, as quantified using $\rho_0$ (Equation~\ref{rho_0}), they argue that it does not change the orbital mixing $\phi$ (Equation~\ref{phi_equation})%; this is in contrast to the effect of chemical doping of JT-inactive ions on the JT-active sites in perovskite systems which does change the orbital mixing~\cite{zhou2011jahn}
. KCuF$_3$ has similar orbital ordering to LaMnO$_3$, except the degeneracy is due to the $d^9$ hole rather than an electron. The variable-pressure crystal structures for KCuF$_3$ are available on ICSD (catalog codes 182849-182857), and are utilised here.

\begin{figure}%[t]
    \includegraphics[scale=0.90]{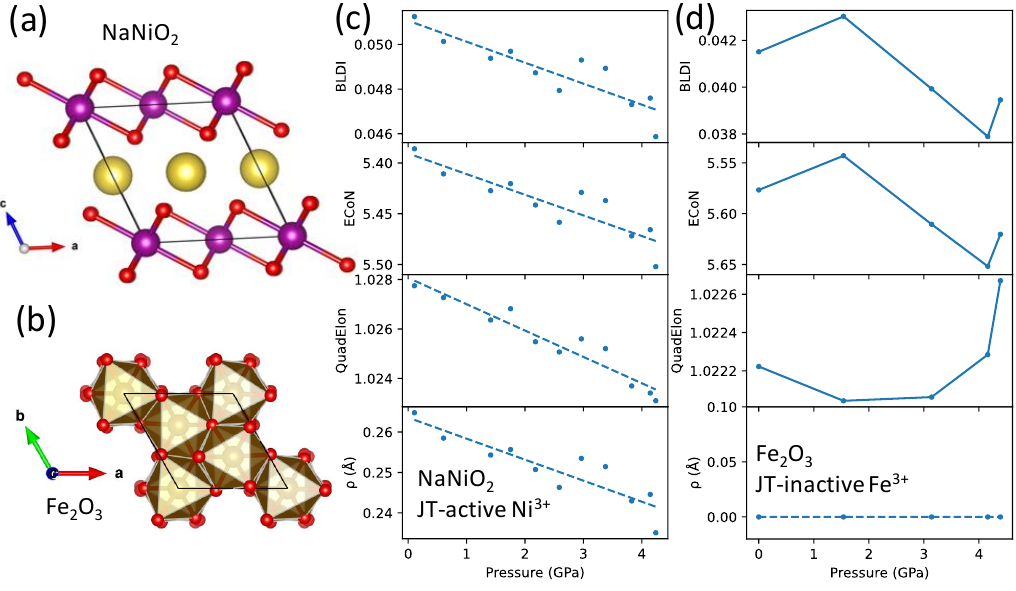}
    \caption{The crystal structures of Jahn-Teller-active $C2/m$
    NaNiO$_2$ and inactive $R\bar{3}c$ Fe$_2$O$_3$ are shown in (a) and (b) respectively. (c) and (d) show a comparison of various metrics for quantifying the degree of Jahn-Teller distortion as a function of pressure, for NiO$_6$ octahedra in NaNiO$_2$ and FeO$_6$ octahedra in Fe$_2$O$_3$ respectively. The parameters subject to comparison are the magnitude $\rho_0$, bond length distortion index, effective coordination number, and quadratic elongation. Dashed lines indicate a linear fit to the data, whereas solid lines connect data points.}
    \label{VanVleck_pressure}
\end{figure}

\begin{figure}%[t]
    \includegraphics[scale=1.00]{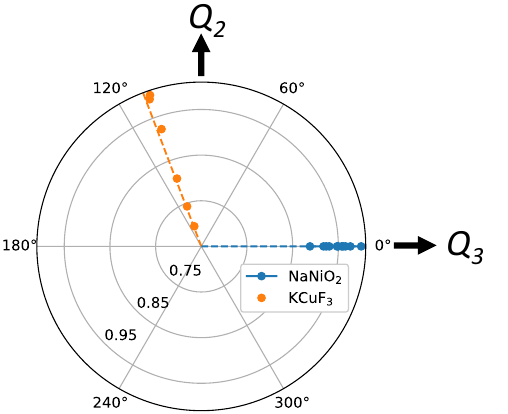}
    \caption{An $E_g(Q_2,Q_3)$ radial plot comparing the pressure-dependence of the $M$O$_6$ ($M$=Ni,Cu) octahedra for KCuF$_3$ and NaNiO$_2$ between 0 and 5\,GPa, where $\rho_0$ is normalised to the value at the lowest measured pressure and the dashed lines represent the average $\phi$ for each material within this pressure range.}
    \label{VanVleck_pressure_polar}
\end{figure}

\begin{figure}%[t]
    \includegraphics[scale=0.90]{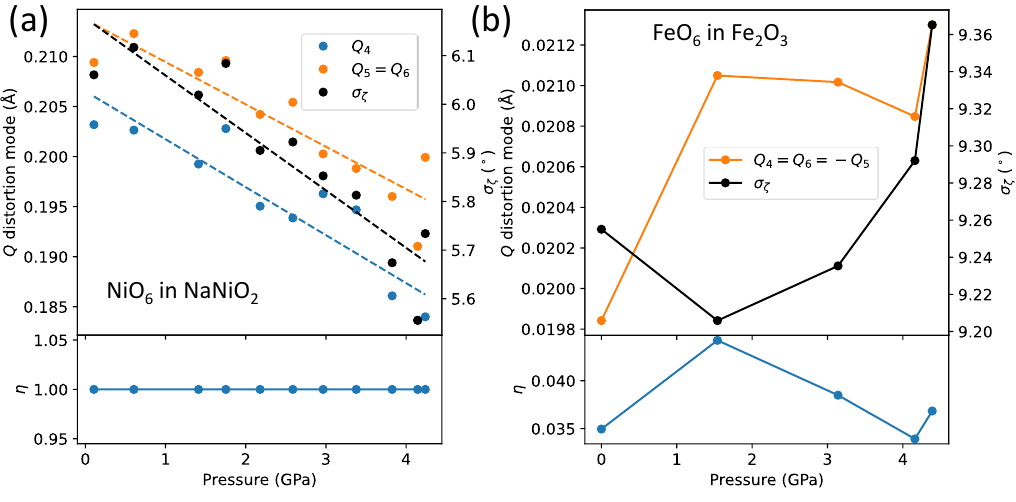}
    \caption{The pressure-dependence of the shear and angular distortion in (a) Jahn-Teller-distorted NiO$_6$ octahedra in NaNiO$_2$ and (b) Jahn-Teller-undistorted FeO$_6$ octahedra in Fe$_2$O$_3$. Shear distortion is represented with the $Q_4$, $Q_5$, and $Q_6$ modes for the octahedra, and angular distortion is represented by bond angle variance. Dashed lines indicate a fitted straight line to the data, whereas solid lines are plotted from point to point. $\eta$ is the angular shear fraction defined in Equation~\ref{eta_equation}. Note that for the NiO$_6$ octahedra, $Q_5=Q_6$, whereas for FeO$_6$ octahedra, $Q_4=Q_6=-Q_5$. For Fe$_2$O$_3$, the average position of the O ligands were taken as the centre of the octahedron.}
    \label{VanVleck_pressure2}
\end{figure}

We previously studied the effect of pressure on the Jahn-Teller distortion in NaNiO$_2$~\cite{nagle2022pressure}, by performing Rietveld refinement~\cite{rietveld1969profile} of neutron diffraction data from the PEARL instrument~\cite{bull2016pearl} at the ISIS Neutron and Muon Source. However, we did not utilise the van Vleck distortion modes, instead quantifying the Jahn-Teller distortion using the bond length distortion index~\cite{baur1974geometry} and the effective coordination number~\cite{hoppe1979effective}. In that study, we found no deviation from the ambient-pressure space group $C2/m$~\cite{dick1997structure,sofin2005new}, shown in Figure~\ref{VanVleck_pressure}(a), for all pressure points at room-temperature up to $\sim$4.5\,GPa. This space group permits only four short\{long\} and two long\{short\} bonds or 6 equal bond lengths, depending on the angle $\beta$, and so throughout the measured pressure range there exists no $Q_2$ character to the Jahn-Teller distortion, consistent with the principle that hydrostatic pressure does not change orbital mixing~\cite{zhou2011jahn}.

Here, we perform a fresh analysis of the variable-pressure octahedral behaviour as a function of pressure at room temperature in NaNiO$_2$ in terms of the $E_g(Q_2,Q_3)$ van Vleck distortion modes. For a reference we sought a material which does not exhibit a first-order Jahn-Teller distortion but does exhibit bond length distortion; for this purpose, we selected Fe$_2$O$_3$, the pressure dependence of which was previously studied in \citename{finger1980crystal} \citeyear{finger1980crystal}, and which exhibits bond length distortion due to its face- and edge-sharing octahedral connectivity. Fe$_2$O$_3$ contains high-spin $d^5$ Fe$^{3+}$ cations within octahedra which interact via both face- and edge-sharing interactions. It should be noted that Fe$_2$O$_3$ likely exhibits some very subtle pseudo Jahn-Teller distortion (related to, but distinct from the first Jahn-Teller effect discussed here) on account of the Fe$^{3+}$ ions~\cite{cumby2017ellipsoidal,bersuker2020perovskite}, but this does not impact the discussion in any meaningful way. 

In Figure~\ref{VanVleck_pressure}(c) we compare (for NaNiO$_2$) $\rho_0$ with three other parameters (bond length distortion index, quadratic elongation, and effective coordination number) which are often used to parametrise the magnitude of the Jahn-Teller distortion. The trend for each is near identical, although the magnitudes differ greatly, indicating that each is a reasonable parameter for quantifying the magnitude of the Jahn-Teller distortion. This can be compared to Figure~\ref{VanVleck_pressure}(e) which shows the same parameters for the Jahn-Teller-undistorted FeO$_6$ octahedra in Fe$_2$O$_3$, where it can be seen that $\rho_0$ remains approximately at zero throughout the measured pressure range, despite a high level of bond length distortion as represented by the bond length distortion index, effective coordination number, and quadratic elongation (a similar plot for KCuF$_3$ can be seen in SI, Figure~S3%\ref{SI_KCuF3_distortion}
). This means that, while these parameters are valid for quantifying the magnitude of Jahn-Teller distortion, they are also sensitive to other kinds of distortion. $\rho_0$ is calculated using $Q_2$ and $Q_3$ which have $E_g$ symmetry, and so $\rho_0$ will only be non-zero for a distortion with $E_g$ symmetry. Thus, it is arguably the ideal choice for parameterising the magnitude of this type of Jahn-Teller distortion. However, while $\rho_0$ is more reliable than the other parameters shown in Figures~\ref{VanVleck_pressure}(c,d) for demonstrating the presence of a Jahn-Teller distortion, it is not always strictly zero for a Jahn-Teller-inactive octahedron, \textcolor{black}{as it will have a non-zero value if the octahedron is distorted with an $e_g$ symmetry}. For example, the NaO$_6$ octahedron in $C2/m$ NaNiO$_2$ has the same symmetry as the NiO$_6$ octahedron, and so exhibits a value of $\rho_0$ between 0.065 and 0.05 within the studied pressure range [Figure~S4 % \ref{SI_NaO6_NaNiO2} 
in Supplementary Information], and Jahn-Teller-inactive FeO$_6$ octahedra in $R$FeO$_3$ perovskites have non-zero $\rho_0$ due to the $E_g$ symmetry of the distorted octahedra, as shown in \citename{zhou2008intrinsic} \citeyear{zhou2008intrinsic}. 

Figure~\ref{VanVleck_pressure_polar} shows a polar plot for the behaviour of NaNiO$_2$ and KCuF$_3$ in the range 0 to 5\,GPa (the measured range for NaNiO$_2$). It can be seen that within this pressure range, the magnitude of the Jahn-Teller distortion decreases far more for KCuF$_3$ than NaNiO$_2$; this reflects the fact that KCuF$_3$ is more compressible, with a bulk modulus 57(1)\,GPa~\cite{zhou2011jahn} compared with 121(2)\,GPa for NaNiO$_2$~\cite{nagle2022pressure}, as obtained by a fit to the third-order Birch-Murnaghan equation-of-state~\cite{birch1947finite}. Within this pressure range we see that $\phi$ does not change with pressure for either material, and that this property is true regardless of whether $\phi$ is or is not a special angle (as in Table~\ref{table_special_angles}), consistent with the interpretation of \citename{zhou2011jahn} \citeyear{zhou2011jahn}. 

Finally, in the previous study~\cite{nagle2022pressure}, we showed using specific O-Ni-O bond angles that pressure reduces the angular distortion for NaNiO$_2$. %\footnote{For KCuF$_3$, octahedral shear is zero throughout the measured range, likely due to the lack of octahedral tilting in this system~\cite{zhou2011jahn}.} 
Here, we show that pressure also reduces the related shear distortion in NaNiO$_2$. This is demonstrated in Figure~\ref{VanVleck_pressure2} where we plot the octahedral shear $Q_4$, $Q_5$, and $Q_6$ modes %\footnote{Note for $C2/m$ NaNiO$_2$, $Q_5=Q_6$, and for $R\bar{3}c$ Fe$_2$O$_3$, $Q_4=Q_6=-Q_5$} 
for NaNiO$_2$ and Fe$_2$O$_3$ against the bond angle standard deviation, $\sigma_\zeta$, defined in Equation~\ref{BAV_equation}. Unlike the AlO$_6$ octahedra in LaAlO$_3$ [Figure~\ref{LaAlO3_figure}], for NiO$_6$ octahedra in NaNiO$_2$ there is no perfect correlation between the shear modes and angular distortion despite $\eta\approx1$, because there is more than one independent shear mode, but we can see that shear distortion and angular distortion are still highly correlated. However, for Fe$_2$O$_3$ the shear fraction $\eta << 1$ and there is no correlation between the shear distortion modes and angular distortion. This difference in behaviour likely arises because the main driver of the change is a continuous decrease in the Jahn-Teller distortion in NaNiO$_2$, as compared to Fe$_2$O$_3$ where positions of the oxygen anions are determined by the reduced degrees of freedom arising from trying to satisfy multiple face- and edge-sharing interactions. This result could only be achieved by calculating the van Vleck modes in a Cartesian coordinate system as outlined in this paper, as opposed to calculating the distortion modes along bond directions, indicating the relevance of calculating the van Vleck modes in this way, and of the shear fraction $\eta$ we propose in this work. 
		
\section{Conclusion}
	
We present \textsc{VanVleckCalculator}, a code package written in \textsc{Python 3} for the calculation of octahedral van Vleck distortion modes. These modes are particularly important for understanding the behaviour of the Jahn-Teller distortion, and we have shown that the parameter $\rho_0$ (which is based on the van Vleck $Q_2$ and $Q_3$ modes) is a more reliable way of quantifying the Jahn-Teller distortion than other oft-used parameters such as the bond length distortion index. 

We show the importance of using a Cartesian set of coordinates for this calculation, instead of calculating the modes along bond directions, as is often done in the literature. This is because calculating the van Vleck distortion modes along bond directions relies on the assumption that there is no angular distortion or octahedral shear, which is often a false assumption and artificially constrains the $Q_4$, $Q_5$, and $Q_6$ modes to be zero. We show that there is value in calculating these later modes, for instance in understanding the effect of octahedral tiling on octahedra in perovskite-like materials. These shear modes will also be useful for parameterising the Jahn-Teller effect when the degeneracy occurs in the $t_{2g}$ orbitals and results in a trigonal distortion, because their symmetry matches the distortion.

We also show that octahedral shear correlates with angular distortion for materials under the influence of tuning parameters such as pressure or temperature where there is a continuously-varying distortion, such as octahedral tilting (as in LaAlO$_3$) or first-order Jahn-Teller distortion (as in NaNiO$_2$). However, there is no correlation when the distortion is due to competing interactions due to face- or edge-sharing octahedra (as in Fe$_2$O$_3$). We propose a new parameter, the shear fraction $\eta$ (defined in Equation~\ref{eta_equation}), which can be used to predict whether there will be correlation between octahedral shear modes and angular distortion.

     % Appendices appear after the main body of the text. They are prefixed by
     % a single \appendix declaration, and are then structured just like the
     % body text.

     %-------------------------------------------------------------------------
     % The back matter of the paper - acknowledgements and references
     %-------------------------------------------------------------------------

     % Acknowledgements come after the appendices
\vspace{10mm}
\ack{Acknowledgements}

The authors thank Andrew L. Goodwin (University of Oxford) for useful discussions and for sharing the LaMnO$_3$ supercell from \citename {thygesen2017local} \citeyear {thygesen2017local}. 
The authors thank James Cumby (University of Edinburgh) for sharing the crystal structures for LaAlO$_3$ originally published in \citename{hayward2005transformation} \citeyear{hayward2005transformation} and subsequently analysed in \citename{cumby2017ellipsoidal} \citeyear{cumby2017ellipsoidal}. 
The authors \textcolor{black}{acknowledge} comments on earlier drafts of this manuscript from James M. A. Steele, Fiamma Berardi, and Venkateswarlu Daramalla, all at the University of Cambridge.
LNC acknowledges Annalena R. Genreith-Schriever (University of Cambridge) and Ben Tragheim (University of Warwick) for useful discussions.

\vspace{10mm}
\ack{Funding}

LNC acknowledges a scholarship EP/R513180/1 to pursue doctoral research from the UK Engineering and Physical Sciences Research Council (EPSRC). 

\vspace{10mm}
\ack{Graphical software}

Graphs and radial plots were prepared using \textsc{Matplotlib}~\cite{Hunter:2007}. Crystal structures figures were made using \textsc{Vesta-III}~\cite{momma2011vesta}.

%% Note added by Overleaf: If using bibtex, remove the "references" environment above, and uncomment the following line.
\nocite{koccer2019cation}
\nocite{swanson1980polyhedral}
\nocite{halasyamani2004asymmetric}
\referencelist{}

     %-------------------------------------------------------------------------
     % TABLES AND FIGURES SHOULD BE INSERTED AFTER THE MAIN BODY OF THE TEXT
     %-------------------------------------------------------------------------

     % Simple tables should use the tabular environment according to this
     % model

\end{document}